\documentstyle[preprint,aps,eqsecnum]{revtex}
\tightenlines
\begin{document}
\draft
\title{Segment Motion in the Reptation Model \\
       of Polymer Dynamics. II. Simulations}
\author{A.\ Baumg\"artner$^1$, U.\ Ebert$^2$, and L.\ Sch\"afer$^3$} \address{$^1$ Institut f\"ur Festk\"orperforschung and Forum
        Modellierung, Forschungszentrum J\"ulich, 52425 J\"ulich, Germany.}
\address{$^2$ Instituut-Lorentz, Universiteit Leiden, Postbus 9506,         
2300 RA Leiden, the Netherlands,}
\address{$^3$ Fachbereich Physik, Universit\"at Essen,
         45117 Essen, Germany,}
\date{submitted to J.\ Stat.\ Phys.\ on September 18, 1997}
\maketitle
\begin{abstract}

We present simulation data for the motion of a polymer chain through a regular lattice of impenetrable obstacles (Evans-Edwards model). Chain lengths range from $N=20$ to $N=640$, and time up to $10^{7}$ Monte Carlo steps. For $N \geq 160$ we for the central segment find clear $t^{1/4}$-behavior as an intermediate asymptote. The also expected $t^{1/2}$-range is not yet developed. For the end segment also the $t^{1/4}$-behavior is not reached. All these data compare well to our recent analytical evaluation of the reptation model, which shows that for shorter times $(t \alt 10^{4})$ the discreteness of the elementary motion cannot be neglected, whereas for longer times and short chains $(N \alt 100)$ tube renewal plays an essential role also for the central segment. Due to the very broad crossover behavior both the diffusion coefficient and the reptation time within the range of our simulation do not reach the asymptotic power laws predicted by reptation theory. We present results for the center-of-mass motion, showing the expected intermediate $t^{1/2}$-behavior, but again only for very long chains. In addition we show results for the motion of the central segment relative to the center of mass, where in some intermediate range we see the expected increase of the effective power beyond the $t^{1/4}$-law, before saturation sets in. Analysis and simulations agree on defining a new set of criteria as characteristic for reptation of finite chains.

key words: reptation, polymer dynamics, Monte Carlo simulations
\end{abstract}

\section{Introduction}

An understanding of the motion of a chain molecule in a surrounding of impenetrable obstacles is of great interest in the physics of polymer melts or dense solutions as well as for polymers diffusing through gels. With special regard to the latter system De Gennes suggested the reptation model \cite{Z1}. Basic to this model is the observation that the crosslinked structure of the gel for short times restricts the motion of the macromolecule to a tube defined by its initial configuration. The motion proceeds by curvilinear diffusion of little wiggles of 'spared length' along the tube. The destruction of the initial tube ('tube renewal') is due to the motion of the chain ends. These may draw back into the tube, thus shortening the tube and creating new wiggles of spared length, or they may unfold and thus destroy spared length, thus prolonging the tube in some random direction. This is the natural thermal motion of a flexible chain between topological constraints. 

Proposed originally for motion through rigid gels, this model extensively has been applied to melts or dense solutions \cite{Z2}. It is generally accepted as a basic scenario of polymer dynamics. A critical examination \cite{Z3}, however, shows that the experimental or computer-experimental evidence for the quantitative reliability of the model is not particularly strong. Searching for the asymptotic power laws predicted by reptation theory one typically finds a range of exponents differing from the predictions, a finding often interpreted as crossover behavior from Rouse-type motion to reptation. This is little more than an excuse, since -, with the exception of Doi's theory of the melt viscosity \cite{Z4} -, no effort seems to have been spent to really work out the predictions of the reptation model beyond (intermediate) asymptotics. Thus, since not only the experiments but also the simulations mostly are concerned with melts or with an immobile, but disordered configuration of obstacles, it is not at all clear whether the results reflect intrinsic properties of reptation or are dominated by other mechanisms like entropic traps in disordered systems or relaxation of the surrounding in melts.

To proceed we need precise knowledge of the quantitative implications of the reptation model in the (computer-) experimental range of time and chain length. We therefore analytically have worked out detailed quantitative predictions of the model, and we have carried through extensive simulations. All our work is concerned with the original reptation scenario: motion of a discrete chain through an ordered lattice of impenetrable obstacles, which confines the internal motion of the chain to a very narrow tube. Our analytical work concentrates on the motion of individual beads. As long as the bead does not leave the original tube, its motion can be calculated rigorously. Tube renewal and thus the motion of the chain ends can be treated only approximately, and we use an approximation inspired by random walk theory. Details and results of our analytical work may be found in the preceeding paper \cite{Z5}.

The present paper is devoted to our simulations. In Sect.\ II we introduce a Monte Carlo model, which first has been proposed by Evans and Edwards \cite{Z6}. We also briefly describe our analytical model as far as needed for some of the arguments to follow, and we discuss the relation among the Monte-Carlo and analytical models. In Sect.\ III we review results of previous simulations of the Evans-Edwards model, compare to corresponding results of our simulations, and discuss the relevant time scales. Sects.\ IV and V are devoted to a detailed comparison with our theory, where in Sect.\ IV we treat the motion of the central bead inside the tube, and in Sect.\ V we are concerned with tube renewal. Quantities involving the center-of-mass, for which at present we have no new analytical results, are discussed in Sect.\ VI. Sect.\ VII summarizes our findings. Preliminary results of our analytical work and our simulations have been published in \cite{Z7}. 

\section{Models}
\subsection{Monte Carlo model}

The Evans-Edwards model \cite{Z6} considers the chain configuration as a random walk of $N^{(MC)}-1$ steps ('segments') on a cubic lattice. The lattice constant $\ell_{0}$ henceforth is taken as the unit of length: $\ell_{0} = 1$. The configuration is fixed by giving the positions $\{{\bf r}_{1},\ldots,{\bf r}_{N^{(MC)}}\}$ of the endpoints of all segments ('beads'). The length of segment $j$ equals $\ell_{0}$, by construction, $|{\bf r}_{j+1} - {\bf r}_{j}| = \ell_{0}$. The obstacles form a second cubic lattice, of lattice constant $m \cdot \ell_{0}$, placed such that its lattice points coincide with centers of the cells of the first lattice. The edges of this lattice are considered as impenetrable. Ref. \cite{Z6} uses $m = 1,2,\ldots,10$, but we consider only $m=1$, thus taking the tube as narrow as possible. This should show reptation in clearest form. As illustrated in Fig.\ 1, this model eliminates all kink-type motions of the chain and leaves only hairpins, i.e., subsequent segments of opposite direction: ${\bf r}_{j+1} - {\bf r}_{j} = - ({\bf r}_{j} - {\bf r}_{j-1})$, free to move. Of course also the end beads can move freely. 

Clearly with regard to the static properties this model is identical to a simple random-walk chain. With our choice of the narrowest tube, $m=1$, also the dynamics is most simple. The obstacle lattice comes into play only implicitly in restricting the motion to that of hairpins and chain ends. We start with an initial random walk configuration of the chain. In one elementary step we randomly choose one bead. If it happens to be the tip of a hairpin or a chain end, it is moved with probability $1/6$ to one of its 6 possible positions (including its original position, of course). This completes the elementary move. Monte Carlo time $t^{(MC)}$ is measured in (on the average) one attempted move per bead. The simulations extended to $t^{(MC)} = 10^{8}$, and chains of lengths $N^{(MC)} = 20,40,80,160,320,640$ were used. We measured in each run correlations over time intervals
 $t_{1}^{(MC)} - t_{0}^{(MC)} \leq 10^{7}$, averaging over $t_{0}^{(MC)}$ ('moving average'). In addition the data were averaged over up to $40$ independent runs. This is important in particular for the longer chains, where the equilibration time $T_{2}^{(MC)}$ of the hairpins comes close to the total time of the run (see Sect.\ 3.2).

For the longest chains $(N = 320,640)$ and largest times $(t^{(MC)} \approx 10^{7})$ the standard deviation of our data reaches 6 \%. Due to the moving time average it rapidly decreases with decreasing chain length and time, being less than 3 \% for $t^{(MC)} \alt 10^{5}$, for all chain lengths. 

\subsection{Analytical model}

We use a version of De Gennes' reptation model \cite{Z1,Z5}, discrete in time and space. The tube is taken as a chain of $N$ segments, connecting beads numbered $0,1,\ldots N$. Particles, each representing a spared length $\ell_{s}$, are sitting on the beads of that chain. The average density of these particles is $\rho_{0}$. The particles randomly and independently hop among neighbouring beads, with hopping probability $p$. They do not interact, so that a given particle does not feel the presence of the others. If a particle moves over a bead, it drags it along and displaces it by a distance $\ell_{s}$ along the tube. (For hairpin motion this is illustrated in Fig.\ 1 b of \cite{Z7}.) Reaching a chain end (bead $0$ or $N$), a particle is absorbed by a virtual reservoir. These reservoirs also randomly emit particles at a rate adjusted to pertain the average density $\rho_{0}$. They serve to ease the analysis of the in principle grand canonical problem.  

To establish the connection to the physical motion of the beads we note that the displacement along the tube of some bead $j$ within time interval $t$ is given by $\ell_{s} |n(j,t)|$, where $n(j,t)$ gives the number of particles having passed bead $j$ from one direction, subtracted by the number of particles coming from the other direction. Since the tube conformation itself is a random walk in space, bead $j$ in space has moved an average distance
\begin{eqnarray}
g_{1}(j,N,t) &=& \left\langle \overline{({\bf r}_{j}(t) - {\bf r}_{j}(0))^{2}} \right\rangle \nonumber \\
&=& \ell_{s} \:\overline{|n(j,t)|} \: \: \:.
\end{eqnarray}
(Cf. Eq.\ (I 2.8); in the sequel ref. \cite{Z5} will be refered to as I.) In Eq.\ (2.1) the pointed brackets denote the average over the chain configurations, and the bar indicates the average over particle diffusion. 

Eq.\ (2.1) holds as long as bead $j$ stays in the initial tube. Tube renewal is driven by the emission and absorption of particles by the reservoir. Emission of a particle shortens the tube by $\ell_{s}$ at the end considered. Thus within time interval $t$ the tube from end zero is destroyed up to the bead $j_{<} = \ell_{s} n_{max}(t)$, where $(- n_{max}(t))$ is the largest negative fluctuation in the occupation number of reservoir $0$ within time interval $t$. Since at time $t$ the tube on the average has been shortened by $\ell_{s} n_{max}(t)$ steps and then rebuilt by another $\ell_{s} n_{max}(t)$ randomly chosen steps, we find for the motion of the endsegment (cf. Eq.\ (I 2.12))
\begin{equation}
g_{1}(0,N,t) = 2 \ell_{s}\:\overline{n_{max}(t)} \: \: \:,
\end{equation}
valid for times $t$ smaller than the tube renewal time $T_{3}$.
Combining these considerations we find the somewhat complicated expression (I 2.13) for the motion of an arbitrary bead including tube renewal effects.

All these are exact expressions within the frame of our model, holding as long as the original tube is not destroyed completely. It turns out that Eq.\ (2.1) can be evaluated rigorously, whereas Eq.\ (2.2) as well as the tube renewal effects on the motion of an arbitrary bead can be handled only approximately. The stochastic process $n(0,t)$, giving the occupation of reservoir $0$, is correlated by the fact that a particle emitted may be reabsorbed at some later time, the decay time of the correlation being given by the time $T_{2}$ a particle needs to diffuse over the whole chain. This correlation renders an exact evaluation of $\overline{n_{max}(t)}$ impossible. As explained in detail in I, sect. V we evaluate $\overline{n_{max}(t)}$ in a 'mean hopping rate' approximation, calculating the contribution to $\overline{n_{max}(t)}$ of a time step $s,\:0 < s \leq t$ as the contribution of an uncorrelated process with properly adjusted hopping rate. In a similar spirit we have constructed an approximation for the tube renewal effects on arbitrary beads. Our explicit expressions for $g_{1}(j,N,t)$ will be recalled later in the context of data analysis. 

\subsection{Relation among the models}

Obviously the particles of the theoretical model roughly correspond to the hairpins, and for long chains and large time, where hopefully the influence of the microstructure is negligible, we expect both models to yield identical results. In practice, however, it is not clear whether the experiments reach such a universal regime, and a more detailed discussion of the relation among the models is appropriate.

We first consider the chain lengths. The endbeads of the MC-chain correspond to the particle reservoirs, and a hairpin absorbs two beads. A hairpin thus effectively walks along a chain of $N^{(MC)}-4$ interior beads. This must be compared to the theoretical model, where a particle hops along a chain of $N-1$ interior beads. Thus we should identify
\begin{equation}
N = N^{(MC)} - 3 \: \: \:.
\end{equation}
For the shorter chains $(N^{(MC)} \lesssim 100)$, this correction cannot be neglected. 

Identifying hairpins and particles we should take the spared length $\ell_{s} = 2$. The density $\rho_{o}$ is less well defined. For the simple random-walk type MC-chain it is not hard to determine the full statistics of the side-branches, i.e., of tree-like structures in which each lattice bond is occupied by an even number of segments. In the limit of long chains the average total number of segments in such side branches amounts to \cite{Z8} $N^{(MC)}/3$, whereas the average number of simple hairpins (of two segments each) tends to $N^{(MC)}/9$. Thus the remaining $N^{(MC)}/9$ segments are contained in larger side-branches, which can be seen as a result of a fusion of simple hairpins. In the particle picture this corresponds to an interaction of particles sitting on the same bead, and with this interpretation we should choose $\rho_{o} \approx 1/6$. However, also other complications must be noted: For a hairpin lying on the chain like in the right part of Fig.\ 1, the separation of the configuration into hairpin and backbone of the chain is not unique, this configuration in fact showing two mobile points. Thus these considerations suggest an order of magnitude for $\ell_{s}, \rho_{o}$, rather than giving precise values. Taking $\ell_{s},\rho_{o}$ as fit parameters, we in Sect.~4 will find that the data rather precisely determine the combination
\begin{equation}
\ell_{s}^{2} \rho_{o} = 1.23 
\end{equation}
which, as shown in I, sect. IV is the only combination relevant in the universal large time regime. With $\ell_{s}^{2} \rho_{o}$ fixed, a range of values $0.15 \alt \rho_{o} \alt 0.3$ yields equivalent fits, and by convention explained below we choose
\begin{equation}
\rho_{o} = 0.22 \: \: \:,
\end{equation}
leading to
\begin{equation}
\ell_{s} = 2.364 \: \: \:.
\end{equation}
These parameters are of the expected order of magnitude, but they also show that the identification of particles and 'free' hairpins should not be taken to literally. Rather the hairpin motion is renormalized by interaction effects, the particles representing `quasi-hairpins'.

We finally consider the relation among the time scales. The theoretical results, considered as function of 
\begin{equation}
\hat{t} = p t \: \: \:,
\end{equation}
for $p \cdot t \gtrsim 1$ essentially are independent of the hopping rate $p$. We by convention take $p = 1/5$, and we henceforth always will use the variable $\hat{t}$. The relation among $\hat{t}$ and $t^{(MC)}$ defines the time scale $\tau$:
\begin{equation}
\hat{t} = \tau\:t^{(MC)} \: \: \:.
\end{equation}
A fit to experiment (see Sect.~4) fairly precisely fixes $\tau$ at a value 
\begin{equation}
\tau = 6.092 \cdot 10^{-2}
\end{equation}
Thus about 17 MC moves correspond to the displacement of a particle by one step. This again is a reasonable result, since following the microscopic motion of a hairpin we may estimate that on the average of the order of 10 moves are needed for a hairpin to jump from one segment to the next. 

Having discussed the relation among the parameters of the theoretical and the MC-model we still need to consider the measured quantity
\begin{eqnarray*}
g_{1}^{(MC)}(j,N^{(MC)},t^{(MC)}) &=& \left\langle ({\bf r}_{j}(t^{(MC)}) - {\bf r}_{j}(0))^{2} \right\rangle \: \: \:.
\end{eqnarray*}
Let $j$ be some interior bead of the MC-chain. With probability $\rho_{H} = 1/9$ it sits in the tip of a simple hairpin, a configuration which in the theoretical model effectively is projected down to the base of the hairpin. Taking into account only simple hairpins we thus find for the relation of $g_{1}^{(MC)}$ to the $g_{1}$ of the analytical model
\begin{eqnarray*}
g_{1}^{(MC)}(j,N^{(MC)},t^{(MC)}) &=& (1 - \rho_{H})^{2} g_{1}(j-1,N^{(MC)} - 3,t)
\\
&+& 2 \rho_{H} (1 - \rho_{H}) \left[ g_{1} (j-1, N^{(MC)} - 3,t) + 1 \right]
\\
&+& \rho_{H}^{2} \left[ g_{1} (j-1, N^{(MC)} - 3,t) + 2 \right] \: \: \:,
\end{eqnarray*}
where we took into account the relation among $N^{(MC)}$ and $N$ as well as the different counting of the beads. This shows that $g_{1}^{(MC)}$ and $g_{1}$ differ by an additive contribution $c_{0}$
\begin{equation}
g_{1}^{(MC)}(j,N^{(MC)},t^{(MC)}) = g_{1} (j-1,N^{(MC)} - 3,t) + c_{0}
\: \: \:,
\end{equation}
where the simple-hairpin contribution to $c_{0}$ is found as 
\begin{equation}
c_{0} = 2 \rho_{H} = 2/9 \: \: \:.
\end{equation}
More complicated side branches will contribute also, but we observe (see Sect.~4) that reasonable changes of $c_{0}$ in fitting to the data can be compensated by readjusting $\rho_{0}$. We thus by convention choose the value (2.11), which then fixes $\rho_{0}$ to the value (2.5). We have checked that for microscopic times $10 \lesssim t^{(MC)} \lesssim 50$, $g_{1}^{(MC)}$ is well represented as 
\begin{eqnarray*}
g_{1}^{(MC)} = \frac{2}{9} + const\:\cdot (t^{(MC)})^{x}, \: \: \: x \approx 1/2 \: \: \:,
\end{eqnarray*}
so that this choice of $c_{0}$ is well justified.

For the endsegments the correction is more important. These in a single move will jump a mean squared distance
\begin{equation}
c_{1} = 2 \: \: \:,
\end{equation}
and this motion is not taken into account in the theoretical model. We thus have 
\begin{equation}
g_{1}^{(MC)}(1,N^{(MC)},t^{(MC)}) = g_{1}(0,N^{(MC)} -3,t) + c_{1} \: \: \:.
\end{equation}
We have checked that the correction $c_{1} = 2$ precisely takes into account the difference in the motion of the end segment and the adjacent interior segment of the Monte-Carlo chain.

Though these corrections are microstructure effects, they cannot simply be ignored. In particular it is important to correct for the endsegment motion, since $g_{1}^{(MC)}(1,N^{(MC)},t^{(MC)})$ reaches values of the order $100$ only for $t^{(MC)} \approx 10^{6}$. 

Besides $g_{1}^{(MC)}$ we also have measured the cubic invariant
\begin{equation}
\hat{g}_{1}^{(MC)}(j,N^{(MC)},t^{(MC)}) = \left[ \left\langle \sum^{3}_{\alpha = 1}\:\left({\bf r}_{j,\alpha}(t^{(MC)}) - {\bf r}_{j,\alpha}(0)\right)^{4}
\right\rangle\right]^{1/2} \: \: \:.
\end{equation}
It is easily checked that this function for an interior bead in our model reduces to the second moment of $n(j,t)$
\begin{equation}
\hat{g}_{1}(j,N,t) = \ell_{s} \left(\overline{n^{2}(j,t)}\right)^{1/2} \: \: \:,
\end{equation}
this relation holding as long as the bead stays in the initial tube. For the endsegment the expression is more complicated and given in appendix B of I    .
Again the relation among $\hat{g}_{1}$ and $\hat{g}_{1}^{(MC)}$ involves microstructure corrections, which, however, are more difficult to estimate and will not be considered.    

\section{A first inspection of the simulation results}
\subsection{Comparison to previous work}

Reptation theory predicts power law behavior 
\begin{equation}
g_{1}(j,N,t) \sim \left\{ \begin{array}{cc}
t^{1/4}, & T_{o} \ll t \ll T_{2} \\
(t/N)^{1/2}, & T_{2} \ll t \ll T_{3} \\
t/N^{2}, & T_{3} \ll t \: \: \:. \end{array} \right.
\end{equation}
Here $T_{0} = O(N^{0})$ is the microscopic time, till the segment motion feels the constraining environment, $T_{2} = O(N^{2})$ is the equilibration time of the internal motion, and $T_{3} = O(N^{3})$ is the reptation time, needed for a complete destruction of the original tube. The last line of Eq.\  (3.1) identifies the diffusion constant of the chain as $D \sim 1/N^{2}$. 

Evans and Edwards \cite{Z6} introduced the above described Monte Carlo model to test these predictions. They used obstacle lattices of spacing $m \cdot \ell_{0}$, $m \leq 10$, and chains of length $N^{(MC)} \leq 80$. The runs seem to extent up to $t^{(MC)} \sim 10^{3}$. Clearly according to present day facilities this is a fairly small scale simulation. Still within the scatter of the data the results for the smallest spacings $m=1,2$ seem to verify the predictions (3.1). In particular the authors observe $D \sim N^{2},T_{3} \sim N^{3}$, as well as $t^{1/4}$-regimes and $t^{1/2}$-regimes for the motion of the central segment. To check these results, Fig.\ 2 shows our data for $g_{1}^{(MC)}(N^{(MC)}/2,N^{(MC)},t^{(MC)})$ in the common doubly-logarithmic representation. As is obvious, a $t^{1/4}$-regime starts around $t^{(MC)} \approx 10^{3}$ and barely is observable for $N^{(MC)} = 80$. It fully is developed only for larger chain lengths. A $t^{1/2}$-regime is not observable for $N^{(MC)} \leq 160$. It may be present for larger chains but its unambiguos identification needs at least a further decade in time. Recall that these data are taken for the obstacle lattice of highest density, $m=1$, as all our data. We conclude that the observation of \cite{Z6} amounts to a misinterpretation of the direct crossover from the initial behavior, which roughly follows a $t^{1/3}$-law, to free diffusion, as seen here for short chains.

Deutsch and Madden \cite{Z9} used the Evans-Edwards model with $m=1$ to measure the diffusion coefficient $D$ by following the center-of-mass motion of the chain:
\begin{equation}
\left\langle \left({\bf R}_{cm}(t^{(MC)}) - {\bf R}_{cm}(0)\right)^{2}\right\rangle 
\stackrel{ t^{(MC)} \rightarrow \infty}{\displaystyle\longrightarrow}
\:D\:t^{(MC)}
\end{equation}
Measuring chains up to length $100$ they found $D \sim \left(N^{(MC)}\right)^{-2.5}$, i.e., a considerably faster decrease than predicted by reptation theory. Our own data for the center-of-mass motion are shown in Fig.\ 3. We clearly can extract the diffusion coefficients up to $N^{(MC)} = 80$. For $N^{(MC)} \geq 320$ only an upper bound $D \leq 0.2 (N^{(MC)})^{-2}$ can be given. Our measured values of $(N^{(MC)})^{2} \cdot D$ are plotted against $(N^{(MC)})^{-1/2}$ in Fig.~4. We also included data extracted from Fig.\ 3 of \cite{Z9}. Clearly the two sets of data are completely consistent. They nicely are fitted by the ansatz
\begin{equation}
D = 0.04 (N^{(MC)})^{-2} \left[ 1 + 50 (N^{(MC)})^{-1/2}\right] \: \: \:,
\end{equation}
this form being motivated by Doi's work \cite{Z4,Z9}. It, however, is clear that the range of chain lengths from $10$ to $160$ in Fig.\ 4 is insufficient to fully justify an ansatz leading to such a large first order correction. Still it shows that with chain lengths that presently can be reached, we are far from extracting the large-$N$ limit of the diffusion coefficient.

\subsection{Time scales}

The reptation time $T_{3}$ gives the time needed to destroy the tube. It may be defined in terms of the decay of the end-to-end vector correlations of the chain \cite{Z1}, but for the present work another definition is more convenient, both theoretically and experimentally. We define $T_{3}$ by the relation
\begin{equation}
\ell_{s}\:n_{max}(T_{3}) = \frac{N}{2} \: \: \:,
\end{equation}
which via Eq.\ (2.2) implies
\begin{eqnarray}
g_{1}(0,N,T_{3}) &=& \left\langle ({\bf r}_{0}(T_{3}) - {\bf r}_{0}(0))^{2}\right\rangle
\nonumber \\
&=& N = R_{e}^{2} \: \: \:,
\end{eqnarray}
where $R_{e}^{2}$ is the mean squared-average end-to-end vector. Thus within time interval $T_{3}$ the endsegment has moved mean squared distance $R_{e}^{2}$. Our simulation data allow for the determination of $T_{3}$ for $N^{(MC)} \leq 160$, and our experimental results together with the theoretical curve are shown in Fig.~5. Being based on an approximation the theory lies about 20 \% above the datapoints. This suggests that our approximation underestimates $n_{max}(t)$. Both theory and data, however, consistently show that it needs chain lengths much larger than $N^{(MC)} = 200$ to approach the asymptotic $N^{3}$ behavior. Indeed, the theoretical asymptote, calculated from Eq.\ (I 5.50), is found as 
\begin{equation}
T_{3}^{(MC)} = 2.62 (N^{(MC)})^{3} \: \: \:,
\end{equation}
suggesting that chain lengths much larger than $10^{3}$ are needed. This is quite consistent with our finding for the diffusion coefficient.

Another important time scale of the model is the internal equilibration time $T_{2}$ of the chain. It gives the time a hairpin needs to diffuse over the whole chain, so that for $t \gg T_{2}$ the motion of all beads is correlated. Theoretically $T_{2}$ can be identified with the Rouse time, i.e , the longest internal relaxation time of a free chain: $p T_{2} = N^{2}/\pi^{2}$. A practicable and precise experimental definition is not easy. Heuristically we could think of the time at which the motion of the central segment bends over from $t^{x_{1}},x_{1} \approx 1/4$ towards $t^{x_{2}},x_{2} \gtrsim 0.5$, but this crossover is quite broad and the power law regimes are poorly defined for shorter chains. We thus here are content with the theoretical definition, valid for long chains:
\begin{equation}
p\:T_{2} = \left(\frac{N}{\pi}\right)^{2}
\end{equation}
Transforming to the Monte Carlo time, we find
\begin{equation}
T_{2}^{(MC)} = 1.66 (N^{(MC)})^{2} \: \: \:,
\end{equation}
where we ignore the difference among $N,N^{(MC)}$. Comparing to Figs.\ 2 or 3 we note that this marks the point, where definite deviations from the initial behavior can be seen. Combining Eqs.\
(3.6),(3.8) we find
\begin{equation}
\frac{T_{3}^{(MC)}}{T_{c}^{(MC)}} = 1.58\:N^{(MC)} \: \: \:.
\end{equation}
Since both, $T_{2}^{(MC)}$ and $T_{3}^{(MC)}$, mark fairly broad crossover regions, the ratio (3.9) must take values of order $10^{3}$ before we can see the intermediate $t^{1/2}$ behavior. (Cf. the second line of Eq.\ 3.1.) Only the longest chain $N=640$ shows a sufficiently large ratio $T_{3}^{(MC)}/T_{2}^{(MC)}$. But then the total MC time $10^{7}$ is not much larger than $T_{2}^{(MC)} = 0.7 \cdot 10^{6}$. Even with the present data we thus have no chance to verify the $t^{1/2}$-law. Qualitative, not quantitative, indications can be found however, as is discussed in sect. VI. These findings are completely consistent with typical results found in the literature for simulations of melts.

\section{Motion inside the tube}
\subsection{Analysis of $g_{1}(\frac{N}{2},N,t)$}

In Eqs.\ (I.4.1), (I.4.2), (I.3.12) we have given the rigorous theoretical result for $g_{1}(j,N,t)$ for motion {\em within the tube}.
To keep track of this condition we here denote this result as $g_{i}(j,N,t)$.
\begin{equation}
g_{i}(j,N,t) = \left(\frac{4}{\pi}\:\ell_{s}^{2} \rho_{0} A_{1}(j,t)\right)^{1/2} \left[ 1 - F_{1} (4 \rho_{0} A_{1} (j,t))\right]
\end{equation}
\begin{eqnarray}
A_{1}(j,t) = \frac{\hat{t}}{N} + \frac{1}{2N} \sum^{N-1}_{k=1} \left[ 1 - \exp \left(- 4 \hat{t}\:\sin^{2} \left(\frac{\pi k}{2N}\right) \right) \right]
\frac{\cos^{2} \left(\frac{\pi k}{N} \left(j + \frac{1}{2}\right) \right)}{ sin^{2} \left(\frac{\pi k}{2N}\right)}
\end{eqnarray}
We use the variable $\hat{t} = p t$, and we somewhat simplified the expression, the simplification being valid for $p t \gtrsim 1$. The correction function $F_{1}(j,t)$ reads (Eq.\ I.3.26)
\begin{equation}
F_{1}(z) = \frac{1}{2 \sqrt{\pi}}\:\int_{0}^{z}\:dx\:x^{-3/2}\:e^{-x} \left( \left(1 - \frac{x}{2}\right)^{-1/2} - 1\right) - \frac{1}{2 \sqrt{\pi}}\:\Gamma \left(- \frac{1}{2},z\right) \: \:.
\end{equation}
It arises from the discreteness of the stochastic variable $n(j,t)$ and is negligible for $z \gtrsim 25$. 

As has been discussed in I, Sect.~4, for long chains and times so large that $F_{1}(z)$ can be ignored, $g_{i}(j,N,t)$ takes the form (cf. Eq.\ I 4.10):
\begin{equation}
g_{i}(j,N,t) = (\ell_{s}^{2} \rho_{0})^{1/2}\: \hat{t}^{1/4} \tilde{g}_{i}\left(\frac{j}{N},\frac{\hat{t}}{N^{2}}\right) \: \: \:,
\end{equation}
\begin{eqnarray}
g_{i}(j,N,t) \longrightarrow \left\{ \begin{array}{ccc}
2 \pi^{-3/4} (\ell_{s}^{2} \rho_{0})^{1/2} \hat{t}^{1/4} &,& \: \: \: \: \: \hat{t}/N^{2} \ll 1 \\
2 \pi^{-1/2} \left(\ell_{s}^{2} \rho_{0} \frac{\hat{t}}{N}\right)^{1/2} &,& \: \: \: \: \: \hat{t}/N^{2} \gg 1  \: \: \:.\end{array} \right.
\end{eqnarray}
It is in this large time region that we determine the nonuniversal parameters. Specifically we get one relation from fitting the $t^{1/4}$-plateau.
\begin{equation}
(t^{(MC)})^{-1/4} g_{i}\left(\frac{N}{2},N,t\right) = 2 \pi^{-3/4} (\ell_{s}^{2} \rho_{0})^{1/2}\:\tau^{-1/4}, \: \:
T_{0}^{(MC)} \ll t^{(MC)} \ll T_{2}^{(MC)}
\end{equation}
Since no $t^{1/2}$-regime properly is reached by the data, we determine $\tau,\ell_{s}^{2} \rho_{0}$ separately by fitting to the crossover at $t \sim T_{2}$, where the $t^{1/4}$-regime terminates. In the fit we exclusively used data for the longest chain: $N^{(MC)} = 640$, so that we have a large region affected neither by initial effects nor by tube renewal. We find the values $\ell_{s}^{2} \rho_{0} = 1.23$ and $\tau = 6.092 \cdot 10^{-2}$ cited in Sect.~ II~C.

With $\ell_{s}^{2} \rho_{0}$ fixed, a variation of $\rho_{0}$ only influences the argument of $F_{1}$ in Eq.\ (4.1). Increasing $\rho_{0}$ we decrease the time range where $F_{1}$ is important. Since $F_{1}(0) = 1$, we thus also increase the initial slope of $g_{i}(j,N,t)$. As mentioned in Sect.II~C, these changes to some extent can be compensated by a change of $c_{0}$ that relates $g_{1}^{(MC)}$ to $g_{1}$. Good fits can be obtained for $0.14 \lesssim c_{0} \lesssim 0.25$, with varying $\rho_{o}$ from $0.3$ to $0.18$. For $\hat{t} \approx 10$, $g_{1}(j,N,t)$ varies by about 15 \%. Larger changes of the parameters lead to a mismatch in the curvature of the theoretical and experimental results. We finally fix all parameters by choosing $c_{0} = 2/9$, leading to $\rho_{0} = 0.22$. 

Fig.\ 6 shows our results for $\log_{10}\left(g_{1}\left(\frac{N}{2},N,t\right)/g_{ass}(t)\right)$ as function of $\log_{10}(\hat{t})$, where $g_{ass}(t)$ gives the intermediate asymptotics defined by the first line of Eq.\ (4.5). This plot focusses on the region of $t^{1/4}$-behavior. As mentioned above, the parameters have been fitted for $N^{(MC)} = 640$, the remaining results involving no further fitting. This plot demonstrates the ability of our reptation model to explain the data. Deviations occuring for shorter chains and large time result from tube renewal, as is obvious from the values of $T_{3}$ indicated by the ends of the full lines. Within a rough approximation such effects will be considered in Sect.~V~B. The discreteness correction $F_{1}(z)$ is visible up to $\hat{t} \approx 10^{3}$, and for the shorter chains this initial range immediately crosses over to behavior dominated by tube renewal, with no indication of an intermediate $t^{1/4}$-plateau. Furthermore, the $t^{1/2}$-regime is not developed within the range of our simulations. 

\subsection{Analysis of $\hat{g}_{1}(\frac{N}{2},N,t)$}

As has been stressed and illustrated earlier \cite{Z7}, the cubic invariant 
$\hat{g}_{1}(\frac{N}{2},N,t)$ (Eq.\ 2.14) should show $t^{1/4}$-behavior even for very small time, with no discreteness correction. For motion inside the tube the theory yields
\begin{equation}
\hat{g}_{i}(j,N,t) = (2 \ell_{s}^{2} \rho_{0} A_{1}(j,t))^{1/2} \: \: \:,
\end{equation}
with $A_{1}(j,t)$ given in Eq.\ (4.2). Fig.\ 7 shows our numerical and analytical results, normalized to $\hat{g}_{ass}(t) = \sqrt{\frac{\pi}{2}}\:g_{ass}(t)$. As expected, the $t^{1/4}$-plateau is seen very clearly. Even for the shortest chain, $N^{(MC)} = 20$, there is an initial range of behavior close to $t^{1/4}$. The plateau-value, however, systematically seems to lie 3-5 \% below the theoretical value. We believe that this might indicate some small effect of the interaction among the hairpins, renormalizing the amplitudes but not changing the power laws. 

\section{Tube renewal effects}
\subsection{Motion of the endsegments}

Tube renewal cannot be treated rigorously, and our approximation, discussed in I, Sect.~V, yields for the motion of the endsegment $j=0$
\begin{equation}
g_{1}(0,N,t) = \sum^{t}_{s=1}\:\frac{1}{s}\:g_{i}(0,N,s)
\end{equation}
where $g_{i}(0,N,s)$ is given in Eq.\ (4.1). In Fig.\ 8 we have plotted $g_{1}(0,N,t)/\hat{t}^{1/4}$ together with our data. Obviously our approximation reproduces the qualitative features of the data and performs not too bad on the quantitative level. It somewhat underestimates the mobility of the endsegment, consistent with our finding for $T_{3}$ (cf. Sect.~III~B), but the relative error decreases with increasing time. This is plausible since its origin lies in a mistreatment of the time-dependent correlations which decay on time scale $T_{2}$. Both theory and experiment agree in exhibiting a very long initial transient, much longer than found for the central segment. Only the longest chains barely reach a $t^{1/4}$-plateau. As mentioned in I, Sect.~V~C, this again is due to the discrete nature of the process. With $\ell_{s}^{2} \rho_{0}$ fixed, the theoretical curves again are fairly insensitive to $\rho_{0}$. We should recall that we have corrected the data by subtracting $c_{1} = 2$ from the motion of the endsegment (cf. Eq.~(2.13)). Without that correction, the data in the initial range would be enhanced somewhat, starting around $.54$ at $\hat{t} = 1$. 

The theory predicts that $g_{1}(0,N,t)/g_{i}(\frac{N}{2},N,t)$ for $T_{0} \ll t \ll T_{2}$ reaches a plateau, which in our approximation is found at a value $4 \sqrt{2}$. For $t \gtrsim T_{2}$ this ratio should decrease again, since all segments start to move coherently. Fig.~9 shows our results for this ratio where for the central segment we in the theoretical result approximately took the tube renewal into account in the form discussed in the next subsection: $g_{1}(\frac{N}{2},N,t) = g_{i}(\frac{N}{2},N,t) + 2 g_{r} (\frac{N}{2},N,t)$. This figure should be quite characteristic for reptation. For $t < T_{2}$ it very clearly shows the enhanced mobility of the endsegment. Note that this ratio for the Rouse model would take a value of 2 in the initial range and drop down to 1 for $t \gg T_{2}$. Here it takes much larger values, coming close to our theoretical estimate. Clearly the asymptotic plateau value can not be estimated from the experiment. This would need much longer chains and longer times. For $t \gtrsim T_{2}$ we see the expected decrease. For $t > T_{3}$ the ratio asymptotically should tend to 1. Also this is seen, but it is clear that it needs times $t \gtrsim 10\: T_{3}$ to approximately attain that limit. (In Fig.\ 9, $T_{3}$ for $N^{(MC)} \lesssim 160$ can be taken from the endpoints of the theoretical curves.)

We also measured the fourth moment $\hat{g}_{1}(0,N,t)$. As we have discussed in appendix B of I, after some initial range the ratio $\hat{g}_{1}(0,N,t)/g_{1}(0,N,t)$ should take values 
\begin{eqnarray}
\frac{\hat{g}_{1}(0,N,t)}{g_{1}(0,N,t)} = \left\{ \begin{array}{cc}
1.0125 & \: \: \: \: \: T_{0} \ll t \ll T_{2} \\
1.085 &  \: \: \: \: \: T_{2} \ll t \ll T_{3} \: \: \:, \end{array} \right.
\end{eqnarray}
where these results again are only approximate. Our data in all the regime $T_{0} \ll t \ll T_{3}$ scatter within bounds $(1.025,1.035)$, with no systematic trend observable. This is completely consistent with the estimate (5.2), if we take into account that within the limitations of our simulation the region $T_{2} \ll t \ll T_{3}$ is not properly developed. 

\subsection{Motion of an arbitrary segment}

The mean squared displacement of an arbitrary segment $j$ not to close to the center of the chain for all $t < T_{3}$ can be written as
\begin{equation}
g_{1}(j,N,t) = g_{i} (j,N,t) + g_{r} (j,N,t) \: \: \:,
\end{equation}
where $g_{i}(j,N,t)$ is the contribution of motion inside the tube as given in Eq.\ (4.1), and $g_{r} (j,N,t)$ is the contribution of the tube renewal, which reaches segment $j$ roughly at a time $T_{R}(j)$ defined by 
\begin{equation}
\ell_{s}\:\overline{n_{max}(T_{R}(j))} = j
\end{equation}
(cf. I, Eq.\ 5.31). In I, Sect.~V~C and appendix C, we have evaluated $g_{r}(j,N,t)$ in a rough approximation, based on the distribution of $n_{max}(t)$ for a simple random walk, with hopping probability adjusted to our theoretical result for $\overline{n_{max}(t)}$. Despite ignoring detailed correlation effects this approximation yields quite reasonable results as is illustrated in Fig.\ 10. We there show the motion of segments $j = 20,40,80$ in a chain of length $N^{(MC)} = 640$. In particular for $j=80$ we see the onset of the initial $t^{1/4}$-plateau, which for $\hat{t} \gtrsim 10^{3.5}$ is destroyed by the hairpins diffusing in from the nearest chain end $(j = 0)$. This in itself would lead to another plateau at relative height $\sqrt{2}$ (cf. I, Sect.~IV), but before this can develop two further effects set in. Hairpins from the other chain end become important, too, (which implies $t \gtrsim T_{2})$, and tube renewal effects are seen. Detailed inspection of Fig.\ 10 shows small but significant deviations among experiment and theory, which must be due to our neglect of correlations in $g_{r}$.

Eq.~(5.3) ceases to be valid close to the center of the chain, since it considers tube renewal only coming from one chain end. However, for $t \lesssim T_{3}$ processes where both chain ends within time interval $t$ have made excursions of length $\approx N/2$ into the tube are rare. This suggests to apply Eq.~(5.3) also to the central segment. The result, which should underestimate the tube renewal effects, is given by the long dashed curves in Fig.\ 6. The short dashed curves follow by weighting $g_{r}(j,N,t)$ by a factor of 2, in trying to take into account the symmetry of the tube renewal process. The results look quite reasonable and clearly demonstrate that the deviations from $g_{i}(\frac{N}{2},N,t)$, as given by the full curves in Fig.\ 6, indeed are due to tube renewal. 

The motion of segments $j=80$ for chains $N^{(MC)} = 160$ and $640$ is compared in Fig.\ 11. It for $N^{(MC)} = 160$ shows the additional mobility due to a superposition of the effects from both ends. 

\section{Quantities involving the center-of-mass}

Our theory at present has not been evaluated for quantities like the center-of-mass motion, which involve two-bead correlations. We thus here only present our numerical results and compare to the qualitative predictions of reptation theory.

\subsection{Center-of-mass motion}

We have measured the correlation function
\begin{equation}
g_{cm}(t) = \left\langle ({\bf R}_{cm}(t) - {\bf R}_{cm}(0))^{2} \right\rangle \: \: \:,
\end{equation}
where ${\bf R}_{cm}(t)$ is the position of the center-of-mass of the chain. Reptation theory predicts the power laws 
\begin{eqnarray}
g_{cm}(t) \sim \left\{ \begin{array}{cc}
t^{1/2}/N & \: \: \: \: \: T_{0} \ll t \ll T_{2} \\
t/N^{2} & \: \: \: \: \: T_{2} \ll t \end{array} \right.
\end{eqnarray}
Note that free diffusion sets in for $t \gg T_{2}$, in contrast to the motion of the internal segment, where diffusional behavior is found only for $t \gg T_{3}$ (cf. Eq.~(3.1)). As shown in Fig.\ 3 our data reach the diffusional regime only for $N^{(MC)} \leq 160$, allowing for an extraction of the diffusion coefficients as discussed in sect. III~A. Here we consider the short time regime. 

Fig.\ 12 shows the combination $g_{cm}(t) \cdot N^{(MC)}/\hat{t}^{1/2}$. For $N^{(MC)} \gtrsim 320$ we indeed find a plateau. To the best of our knowledge this is the first time that the $t^{1/2}$-behavior for the center-of-mass motion has been observed. The plateau seems to approach an asymptotic value close to $1.5$, but the splitting of the curves even in the initial range indicates the existence of sizeable corrections to the $N$-dependence. For shorter chains the
large mobility of the chain ends ruins the $t^{1/2}$-behavior and a glance to Fig.\ 3 shows that effective power laws $g_{cm} \sim t^{x}, \frac{1}{2} < x \lesssim .8$, might be extracted, for $t < T_{2}$. 

\subsection{Motion of the central segment relative to the center-of-mass}

The correlation function $g_{2}(j,N,t)$ defined as 
\begin{equation}
g_{2}(j,N,t) = \left\langle \overline{\Big(\left[{\bf r}_{j}(t) - {\bf R}_{cm}(t)\right] - \left[{\bf r}_{j}(0) - {\bf R}_{cm}(0)\right]\Big)^{2}} \right\rangle
\end{equation}
measures the motion of segment $j$ relative to the center-of-mass. For $t \ll T_{3}$ the center-of-mass moves much slower than any specific segment, and thus
\begin{eqnarray*}
g_{2}(j,N,t) \approx g_{i}(j,N,t), \: \: \: \: \: t \ll T_{3} \: \: \:.
\end{eqnarray*}
For $t \gg T_{3}$, however, $g_{2}(j,N,t)$ saturates at some $j$-dependent value, that for $j = N/2$ equals the mean squared radius of gyration $R_{g}^{2}$, up to correction of order $1/N^{2}$. Reptation theory thus predicts 
\begin{eqnarray}
g_{2} (\frac{N}{2},N,t) \sim \left\{ \begin{array}{cc}
t^{1/4}&, \: \: \: \: \: T_{0} \ll t \ll T_{2} \\
(t/N)^{1/2}&, \: \: \: \: \: T_{2} \ll t \ll T_{3} \\
R_{g}^{2} = N/6&, \: \: \: \: \: T_{3} \ll t \: \: \:. \end{array} \right.
\end{eqnarray}
Specifically in some intermediate range $g_{2}(\frac{N}{2},N,t)$ increases with a larger effective power of $t$ than in the initial range, and this phenomenon here is not mixed up with crossover towards free diffusion. Its observation thus is a clear signal of reptation.

Fig.\ 13 shows our results for $N^{(MC)} = 80,160,640$, normalized to the $t^{1/4}$-plateau $g_{ass}(t)$. The sequence of the three regimes (6.4) for $N^{(MC)} = 80,160$ is clearly seen, the data also saturating at $R_{g}^{2}$, as expected. As also was to be expected, in the intermediate range $T_{2} \lesssim t < T_{3}$ the power law $g_{2} \sim t^{1/2}$ is not fully attained, but still the data in this plot show a pronounced maximum. To our knowledge this is the first time that the intermediate $t^{1/2}$-regime has clearly been identified. In Fig.\ 13 we included theoretical curves for $g_{i} (\frac{N}{2},N,t)$ (Eq.\ 4.1) to check whether $g_{2}(\frac{N}{2},N,t)$ for $t \ll T_{3}$ indeed equals $g_{i}(\frac{N}{2},N,t)$. Taking into account that the data are not measured with very good statistics, being averaged over only $10$ independent runs each, we find a very satisfactory agreement. 

\section{Conclusions}

We have performed extensive simulations of the Evans-Edwards model up to chain lengths $N=640$ and $10^{7}$ Monte Carlo time steps. Our simulation data show all features predicted by reptation theory, in particular:
\begin{itemize}
\item[1)] We find a strong increase in mobility of the endsegment, as compared to the central segment.
\item[2)] The simulations of the motion $g_{2}$ of the central segment relative to the center-of-mass exhibit all three time regimes predicted by reptation, including the intermediate '$t^{1/2}$'-regime.
\item[3)] The crossover time $T_{2}$ to '$t^{1/2}$'-behavior of $g_{2}$ coincides with the crossover time to free diffusion of the center of mass.
\end{itemize} 
These features clearly distinguish reptation from pure Rouse motion or a Rouse model with randomly spaced entropic traps \cite{Z10,Z11}.
\begin{itemize}
\item[4)] We also found the celebrated $t^{1/4}$-law for motion of an inner segment, and the corresponding $t^{1/2}$-law for the center of mass. However, we need chain lengths $N > 100$ and correspondingly long times to reliably identify such asymptotic laws. 
\item[5)] Within the range of our simulations, the asymptotic $N$-dependence predicted for the diffusion coefficient $D$, and the reptation time $T_{3}$ is not yet reached. From our results we estimate that chain lengths larger than $N = 10^{3}$ are needed to come close to asymptotics for these quantities. 
\end{itemize}

Our work shows that strong preasymptotic effects are an inherent feature of reptation. Such effects in fact dominate the chain length and time range covered by our simulations. The crossover regions here in particular cover all the region $T_{2} < t < T_{3}$, masking in $g_{1}(j,N,t)$ the expected $t^{1/2}$ behavior for an inner segment. Also crossover from the initial behavior to the $t^{1/4}$-law is very slow, a feature which we trace back to the discrete character of the basic dynamics. Indeed, this crossover is so slow that no $t^{1/4}$-regime is seen for the endsegment. As Figs.\ 5-11 and 13 show, this is well explained by our theory. We indeed find very good agreement between our simulation data and our analytical evaluation of De Gennes' reptation model, also in the regions where the asymptotic predictions fail. To reach this agreement for the large variety of quantities and the large parameter range considered, we have adjusted the four parameters $\rho_{0} \ell_{s}^{2}, \rho_{0}, c_{0}$ and $\tau = pt/t^{(MC)}$ within the physically reasonable range. Our analytical predictions are exact, as long as tube renewal is negligible. Deviations between theory and simulations are mainly due to our only approximate analytical evaluation of tube renewal.  

Will our numerical and analytical results be stable under a change of the microscopic structure of the system? This is a question of many different facets. It is expected that certain types of time-independent disorder in the surrounding, entropic traps \cite{Z10} in particular, may ruin reptation all together. Also the consequences of relaxation of the surrounding, like in a melt, at present are not well known. Restricting ourselves to the original reptation scenario, i.e., to motion through an ordered array of obstacles, we should consider the effect of excluded volume interactions among the beads of the chain. In I, Sect.~II~C, we have given reasons why we believe this to be basically irrelevant here, changing only the embedding of the tube into real space as well as the time scale. More serious is the fact that both in our simulations and our analytical work we use a very narrow tube. Allowing for more degrees of freedom of the chain per unit spacing of the obstacle lattice, we certainly will increase microstructure corrections related to excursions of the bead considered from the center of the tube, and at the same time the discreteness corrections, playing such an important role in the initial time range within our model, will decrease. These effects to some extent may compensate each other. Indeed, in some preliminary simulation using a spring-and-bead chain in continuous space and obstacles of finite diameter in a regular lattice of wider spacing, we found results closely resembling the initial time range of our model presented here. (With this other model we, however, were unable to reach chain lengths and a time range where reptation predictions like the $t^{1/4}$-law properly are found. Rather we stayed with effective $t^{1/3}$-behavior familiar from previous work. With regard to the range of chain lengths and times this is completely consistent with the present findings.) We thus expect that certainly on the qualitative and presumably also on a semiquantitative level our results stay valid for wider tubes even in the initial time range. Of course a naive rescaling of our results in such a problem involving several time- and length-scales may be questionable. 

We finally should comment on consequences of this work for previous work on polymer motion through more realistic environments. Let us first consider polymer motion through a fixed disordered background of other chains, roughly modelling a gel. There the general folklore tells us that the reptation scenario is valid. To examine this we appeal to ref. \cite{Z12}, where the motion of a long chain $(N = 200)$ was simulated. Fig.\ 6 of that work indeed shows more than one decade of $t^{1/4}$-behavior for $g_{2}(\frac{N}{2},N,t)$, i.e , the motion of the central segment relative to the center-of-mass. However, the characteristic increase of the effective power at the end of the $t^{1/4}$ range, as shown in Fig.\ 13, is missing, this law fairly abruptly ending in saturation. Also $g_{1}$ and $g_{2}$ deviate rapidly, $g_{1}$ reaching only an effective $t^{1/3}$-law. Both these observations are not compatible with our reptation results, and this conclusion is strengthened by a glance to the motion of the center-of-mass, as shown in Fig.\ 7, ref. \cite{Z12}. To our feeling this suggests that the published work shows disorder effects \cite{Z10,Z11,Z13} rather than reptation.

With regard to polymer motion in melts the situation is even less clear. Again considering as example some extensive published work \cite{Z14}, we note that many results shown there resemble our results found for short chains. In particular, the ratio $g_{1}(0,N,t)/g_{1}(\frac{N}{2},N,t)$ reaches values of order 2.7, i.e ,
larger than for a Rouse chain. For $g_{2}$ no tendency towards $t^{1/2}$-behavior is seen, but this might be due to the effective shortness of the chains, reaching only of the order of 6 entanglement lengths. Thus it is not unlikely that these results reflect 'reptational' behavior of a very short effective chain.

Clearly with regard to such more complicated systems much work still has to be done, and we hope to have contributed to this task by clearly exhibiting the quantitative consequences of the reptation model.

\vspace{2cm}

{\Large Acknowledgement}

This work was supported by the Deutsche Forschungsgemeinschaft, SFB `Unordnung und grosse Fluktuationen'. Furthermore financial support of UE by the Dutch research foundation NWO and by the EU-TMR-network `Patterns, Noise and Chaos' is gratefully acknowledged.

\newpage
{\Large Figure captions}

Fig.\ 1 \\
An allowed hairpin-move and a forbidden kink-jump, illustrated for a chain embedded in a square lattice with obstacles in the centers of the cells. Our simulation uses the $3$-dimensional version of this model.\\ 

Fig.\ 2 \\
$\log_{10} {\scriptstyle\left( g_{1}^{(MC)} \left(\frac{N}{2}^{(MC)},N^{(MC)},t^{(MC)}\right) \right)}$ as function of $\log_{10}(t^{(MC)})$ for $N^{(MC)}$ = 20,40,80,160,320,640. The straight lines correspond to power laws $g_{1} \sim t^{1/3},t^{1/4},t^{1/2}$ or $g_{1} \sim t$.\\

Fig.\ 3 \\
$\log_{10} R_{cm}^{2}, R_{cm}^{2} = \left\langle \left({\bf R}_{cm}(t^{(CM)}) - {\bf R}_{cm}(0)\right)^{2} \right\rangle$ as function of $\log_{10}\:t^{(MC)}$. Chain lengths $N^{(MC)} = 20,40,80,160,320,640$. The lines indicate the asymptotic behavior $D t^{(MC)}$ (3.2). The thus determined $D (N^{(MC)})$ is further evaluated in Fig.\ 4.\\

Fig.\ 4 \\
$(N^{(MC)})^{2}\:D$ as function of $(N^{(MC)})^{-1/2}$. Dots: present work, ellipsoids: ref. \cite{Z9}. The straight line represents Eq.\ (3.3).\\

Fig.\ 5 \\
$(N^{(MC)})^{3}\:T_{3}^{(MC)}$ as function of $N^{(MC)}$. The curve gives the theoretical prediction. The broken line gives the theoretical asymptote. Points from our simulations.\\ 

Fig.\ 6 \\
$\log_{10}\left(g_{1}\left(\frac{N}{2},N,t\right)/g_{ass}(t)\right)$ as function of $\log_{10}(\hat{t})$. Points are our data. From left: $N^{(MC)} = 20,40,80,160,320,640$. The full lines give the result (4.1), valid for motion inside the tube. The broken lines approximate tube renewal, as explained in sect. 5.2. All lines, except for $N \geq 320$, end at $T_{3}(N)$. The horizontal broken line gives the intermediate asymptotics, here normalized to 1 by dividing through $g_{ass} = 2 \pi^{-3/4}\:(\ell_{s}^{2} \rho_{0})^{1/2}\:\hat{t}^{1/4}$.\\

Fig.\ 7 \\
As figure 6, but for the fourth moment $\hat{g}_{1}\left(\frac{N}{2},N,t\right)/\hat{g}_{ass}(t)$. Curves are for motion inside the tube.\\

Fig.\ 8 \\
$\log_{10} (g_{1}(0,N,t)/\hat{t}^{1/4})$ as function of $\log_{10}(\hat{t})$. Data for $N^{(MC)} = 20-640$ (from left). The curves are calculated within the approximation (5.1) and end at $T_{3}(N)$.\\

Fig.\ 9 \\
$g_{1}(0,N,t)/g_{1}\left(\frac{N}{2},N,t\right)$ as function of $\log_{10}(\hat{t})$.
Data and theory like in Fig.\ 8. The broken line indicates the theoretical plateau value $4\sqrt{2}$.\\ 

Fig.\ 10 \\
$\log_{10} (g_{1}(j,N,t)/g_{ass}(t))$ as function of $\log_{10}(\hat{t})$ for $j = 20,40,80$, and $N^{(MC)} = 640$. The broken line gives the contribution $g_{i}(j,N,t)$, the full line is our result (5.3).\\ 

Fig.\ 11 \\
$\log_{10} (g_{1}(j,N,t)/g_{ass}(t))$ as function of $\log_{10}(\hat{t})$ for $j=80$ and $N^{(MC)} = 160,640$. Curves give the full theory.\\

Fig.\ 12 \\
$g_{cm}(t) N^{(MC)}/\hat{t}^{1/2}$ as function of $\log_{10}(\hat{t})$ for  $N^{(MC)} = 20,40,80,160,320,640$ (from left). Values of $\hat{T}_{2}(N)$, calculated according to Eq.\ (3.7), are indicated by the arrows.\\

Fig.\ 13 \\
$\log_{10} \left(g_{2}\left(\frac{N}{2},N,t\right)/g_{ass}(t)\right)$ as function of $\log_{10}(\hat{t})$. Chain lengths are indicated. The fat curves give $g_{i}\left(\frac{N}{2},N,t\right)/g_{ass}$. The thin lines represent the asymptotic law $g_{2} = R_{g}^{2}$. The thin broken line illustrates a power-law $g_{2} \sim t^{1/2}$.    
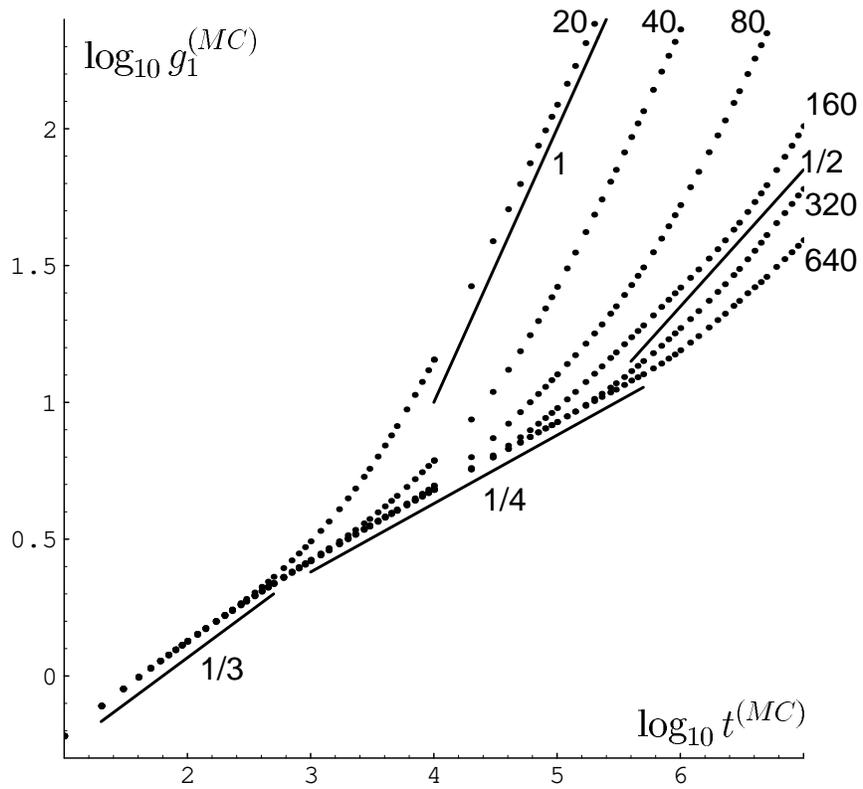
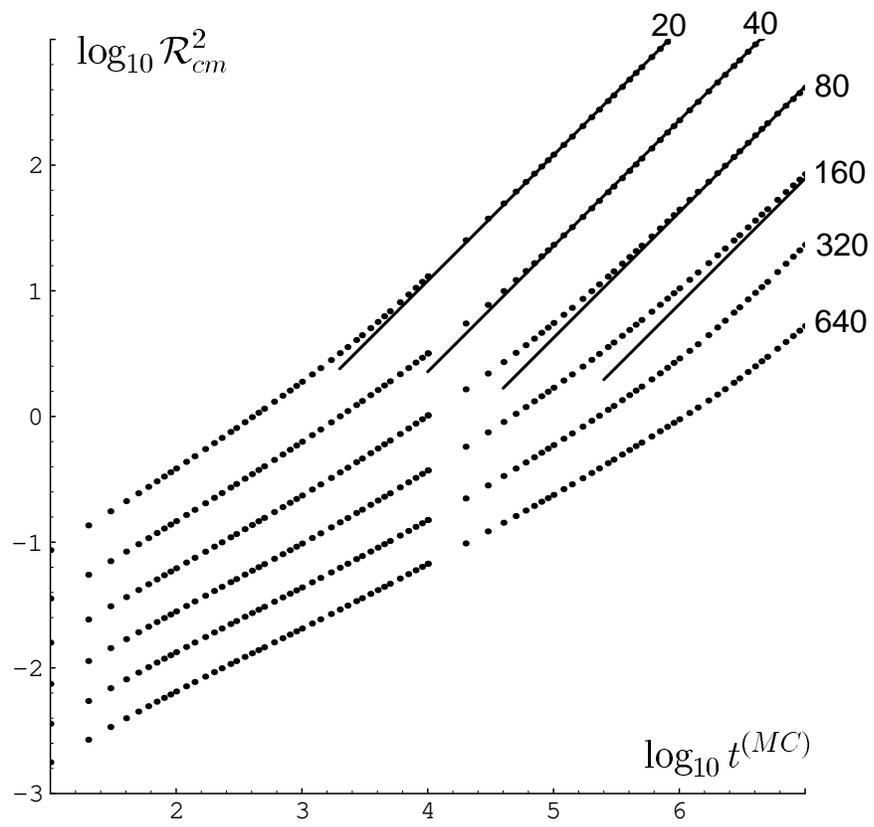
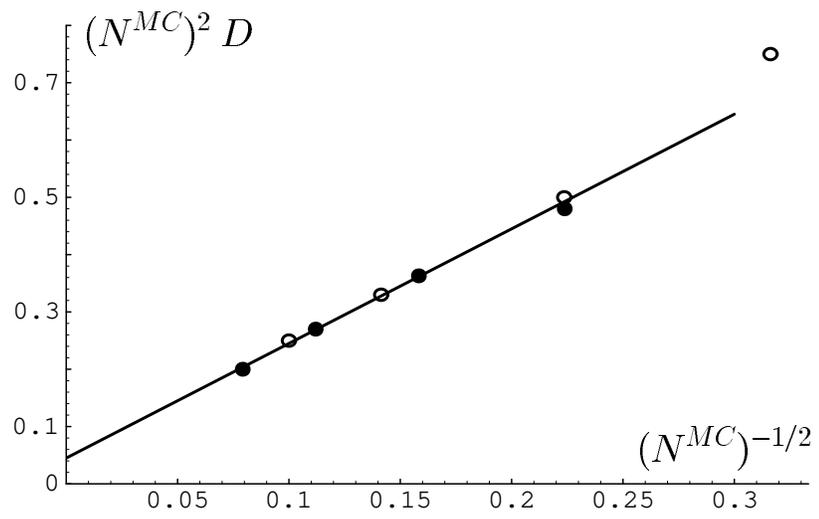
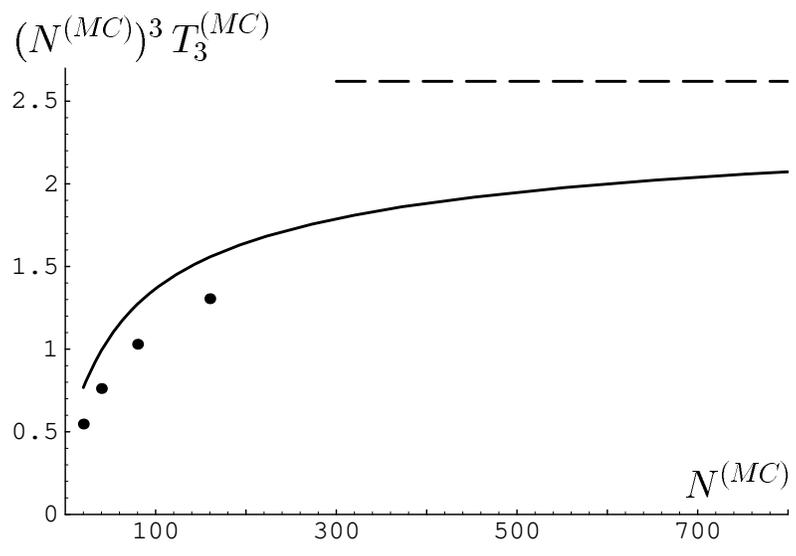
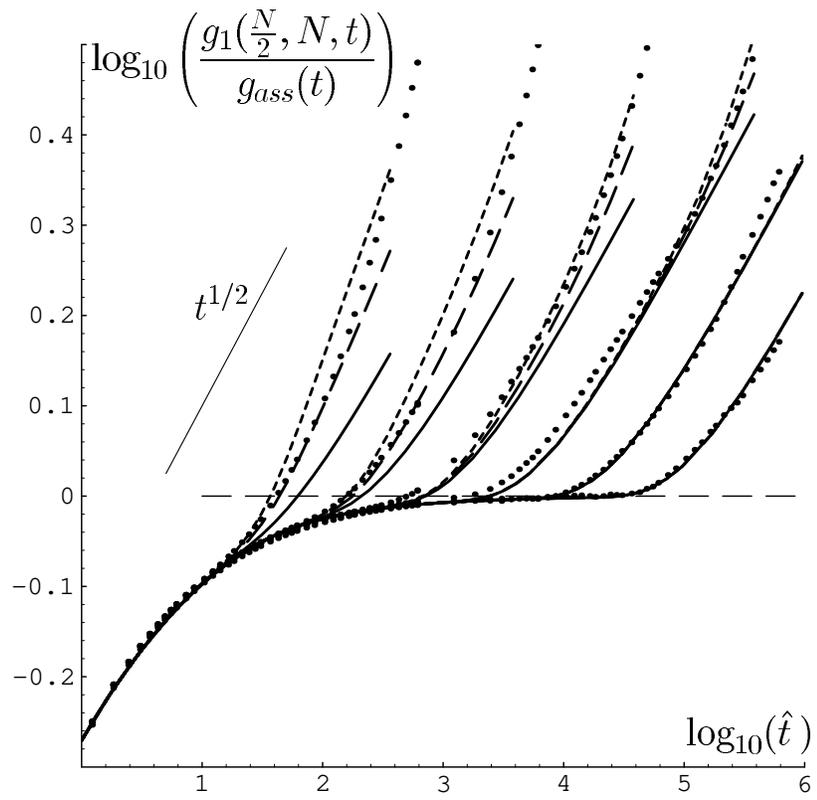
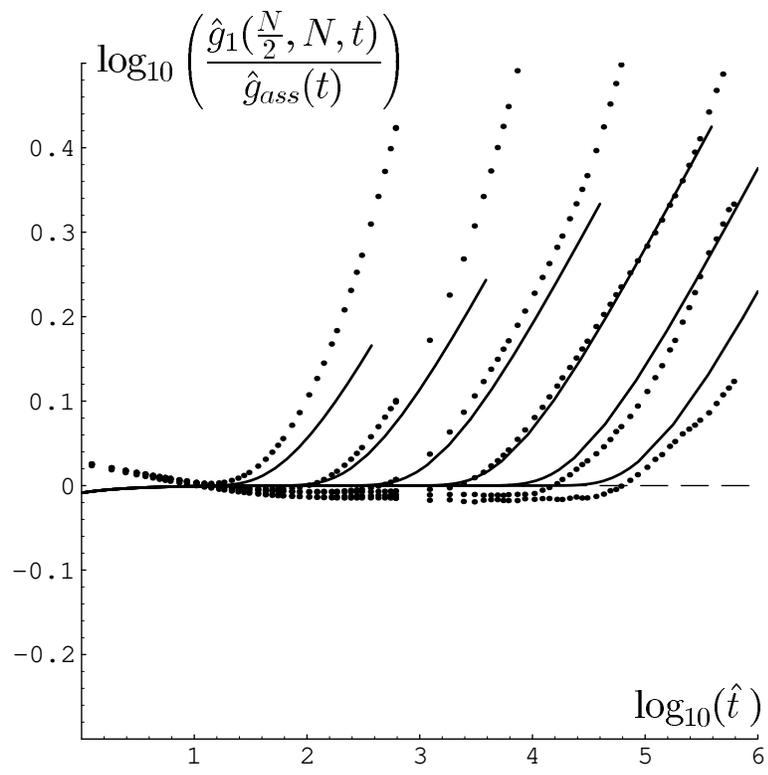
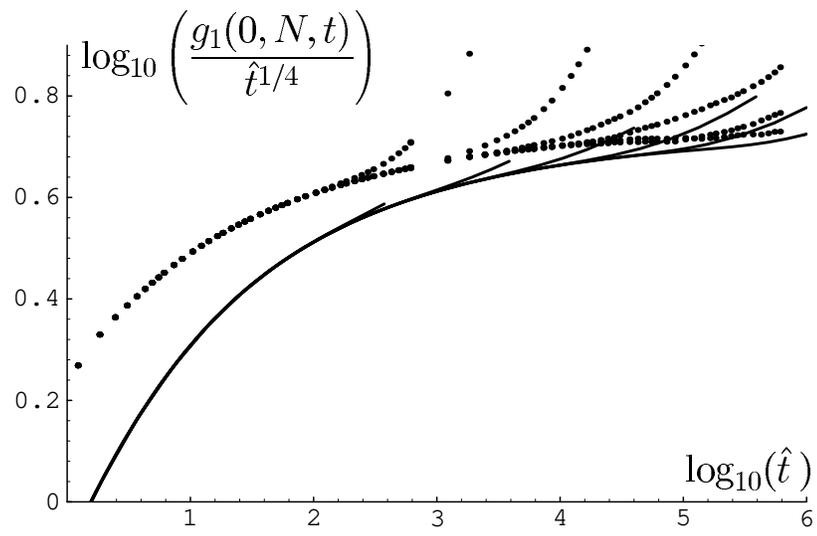
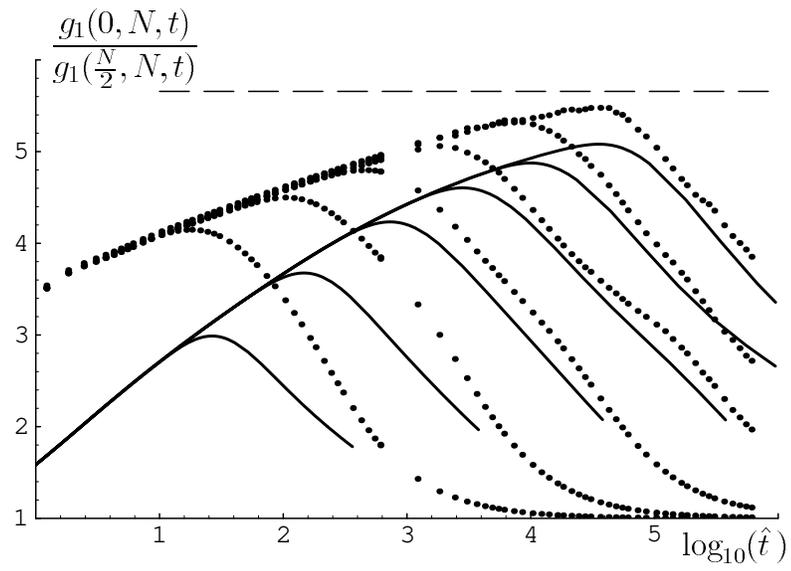
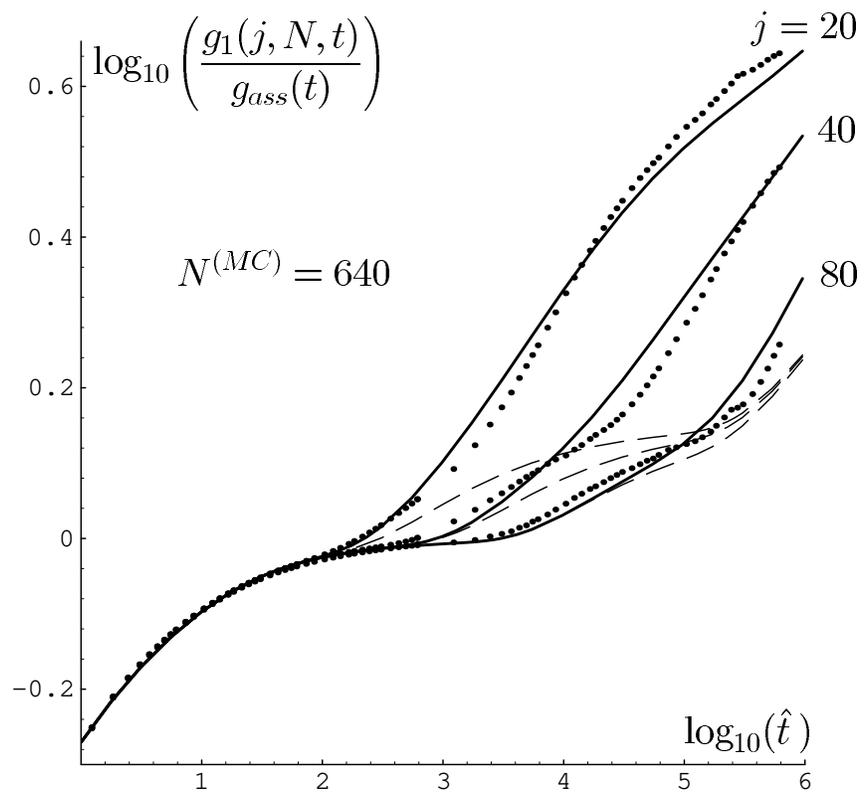
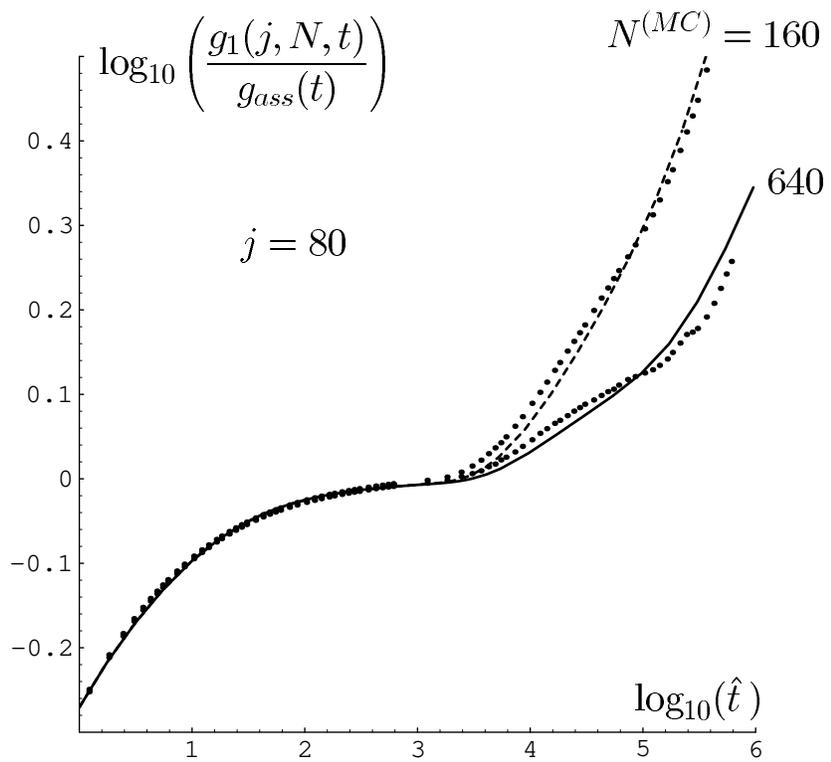
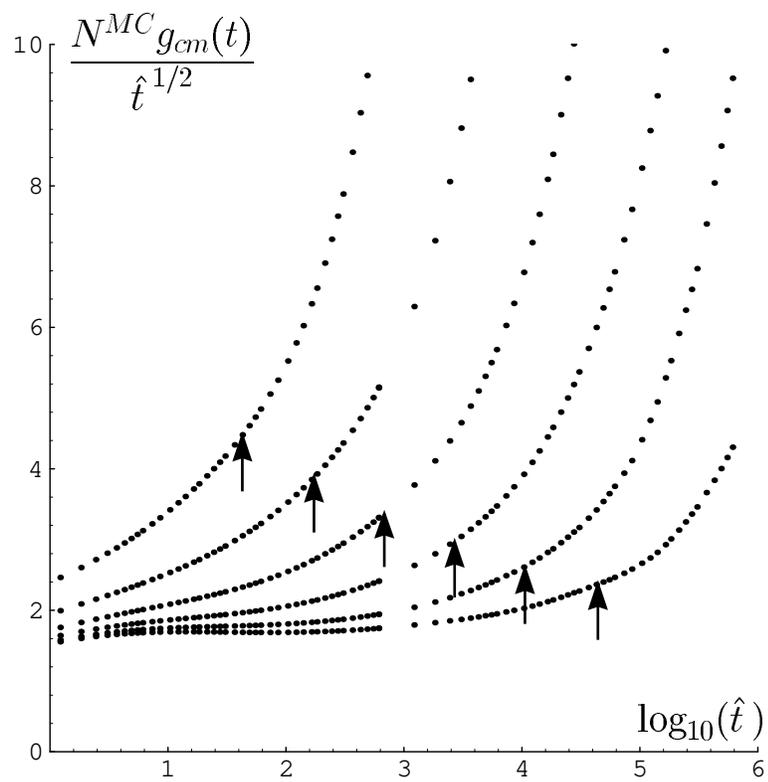
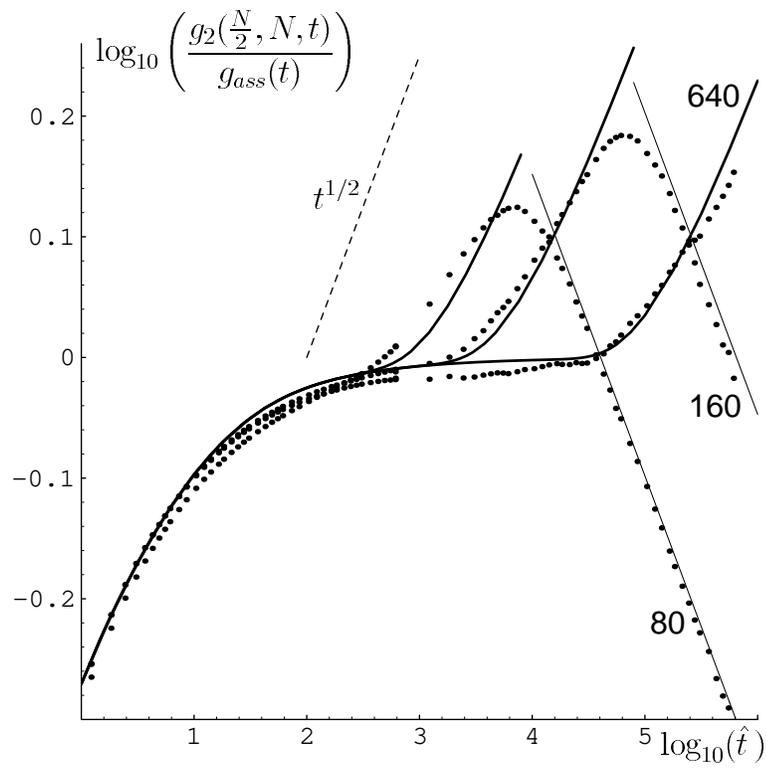
\end{document}